\documentstyle[12pt,epsf,epsfig]{article}
\begin{document}
\begin{titlepage}
\pagestyle{empty}
\baselineskip=21pt
\vspace*{0.20in}
\begin{center}
{\large{\bf Anomaly distribution in quasar magnitudes: a test of lensing 
by an hypothetic Supergiant Molecular Cloud in the Galactic halo}}
\end{center}
\begin{center}
\vskip 0.05in
{\bf Edmond Giraud}$^{1,2}$
\vspace*{0.05in}

{\it
$^1$ {Laboratoire Univers et Particules,
              UMR5299 CNRS-In2p3/Montpellier University, F-34095 Montpellier}\\
$^2$ {Laboratoire d'Astrophysique de Marseille,
              UMR7236 CNRS-INSU/Aix-Marseille University, F-13388 Marseille}\\
}
\vspace*{0.20in}
{\bf Abstract}
\end{center}
\baselineskip=18pt \noindent
   

An anomaly in the distribution of quasar magnitudes based on the 
SDSS survey, has been recently reported by Longo (2012). The angular size
of this anomaly is of the order of $\rm \pm 15^o$ on the sky.  A low surface 
brightness smooth structure in $\gamma$-rays, coincides with the sky location 
and extent of the quasar anomaly, and is close to  the Northern component of 
a pair of $\gamma$-ray bubbles discovered in the \sl Fermi Gamma-ray Space 
Telescope \rm survey. Molecular clouds are thought to be illuminated by 
cosmic rays. A large fraction of molecular gas in the Galaxy is under the 
form of cold $\rm H_2$, and may be a significant component of dark matter.
I test the hypothesis that the magnitude anomaly in the quasar distribution,
is due to a lensing effect by an hypothetic Supergiant Molecular Cloud (SGMC) 
in the Galactic halo.
A series of grid lens models are built by assuming firstly that a SGMC is
a lattice with clumps of $\rm 10^{-3}~M_\odot$, 10 AU in size, and assuming 
various filling factors of the cloud, and secondly a fractal structure. Local 
amplifications are calculated for these lenses  by using the
public software LensTool, and the single plane approximation.
A complex network of caustics due to the clumpy structure is present. Our best 
single plane lens model capable of explaining Longo's effect, \sl at least in 
sparse regions, \rm requires a mass $\rm (1.5-4.1) \times 10^{10} ~M_\odot$ 
within $\rm 8.7 \times 8.7 \times (5-8.6)~ kpc^3$ at a lens plane distance 
of 20 kpc. It is constructed from a molecular cloud building block of 
$5 \times 10^5~M_\odot$ within a scale of 30 pc expanded by fractal scaling 
with dimension $D = 1.8-2$ up to 5-8.6 kpc for the SGMC.
If such a Supergiant Molecular Cloud were demonstrated, 
it might be part of a lens explanation for the luminous anomaly discovered in 
quasars and in red galaxies. The mass budget may be varied by changing the 
cloud depth and the fractal dimension.

\vfill
\vskip 0.15in
\end{titlepage}
\baselineskip=18pt

\section{Introduction}

An anomaly in the distribution of quasar magnitudes based on the 
SDSS survey, was recently reported by Longo (2012). The effect, which is an
enhancement of the order of 0.2 mag, is statistically extremely significant 
in amplitude, and rather well defined in angular size. A first hint of
this anomaly was in fact visible in quasar data as early as three decades ago
(Giraud \& Vigier 1983). Abate \& Feldman (2012) reported
a similar enhancement in the distribution of SDSS luminous red galaxies.
Because the luminous anomaly does not seem to depend on color, a possible physical 
explanation is gravitational lensing. Nevertheless, if the lens is at cosmological distance, 
its huge angular extent translates into a lens 
radius $\sim \rm 350~Mpc$, and its critical surface density implies a total
mass of the order of $\rm 10^{21}~M_\odot$, which is barely consistent with 
$\Lambda$CDM cosmology. A major reason to test a lensing hypothesis with a lens at
\sl short \rm  distance from the observer, possibly in the Galactic halo, is the large 
angular extent of the effect combined with its low amplitude. 

Two large $\gamma$-ray low surface brightness bubbles, extending $50^o$ above
and below the Galactic center were discovered by Su et al (2010) in \sl Fermi Gamma-ray Space Telescope
\rm survey.  These bubbles are spatially correlated with a microwave excess 
in WMAP. Both the \sl Fermi \rm bubbles and the WMAP haze can be 
explained by synchrotron and inverse-Compton from a hard electron population (Dobler et al 2010). In addition to the bubbles in the \sl Fermi \rm map, there is a single large 
faint structure East of the Northern bubble ($l \approx -60^o$, $b \approx 60^o$), that is 
at the location of the quasar anomaly. Any molecular cloud, if present in this faint structure, would be illuminated by cosmic rays escaping from the Northern bubble. 
Cosmic ray/nucleon interactions in molecular clouds were discussed by e.q. Aharonian \& Atoyan (1996), de Paolis et al (1999),  Kalberla et al (1999), Sciama (1999), Shchekinov (2000). The spectra of $\gamma$-rays arising from molecular clouds illuminated by cosmic rays were recently re-examined by Ohira et al (2011). The angular extension of the 
$\gamma$-ray structure is consistent with that of the quasar luminosity anomaly. 
We make the working hypothesis that $\gamma$-rays coming from the faint 
structure trace a SuperGiant cold Molecular Cloud (SGMC) in the Galactic halo.

A large fraction of molecular gas in the Galaxy is under the form of 
cold $\rm H_2$, and may be a significant component of dark matter (Pfenniger, Combes, 
\& Martinet 1994). The variation of universal rotation curves (Persic, Salucci, \& Stel 1996) with 
mass in spiral galaxies may be interpreted as an increasing fraction of baryonic dark matter 
with decreasing mass of galaxies (Giraud 2000a,b; 2001). The fractal structure of molecular gas clouds, 
their formation, thermodynamics and stability were intensively studied (Scalo 1985; Henriksen 1991; 
Falgarone, Phillips, \& Walker 1991; Walker \& Wardle 1998; Irwin, Widraw, \& English 2000). 
The existence of small dark clumps of gas in the Galaxy was invoked to
explain Extreme Scattering Events of quasars (Fielder et al 1987, 1994).
The duration of such events suggests sizes and masses of individual cloudlets of the order ~10 AU and
~$\rm 10^{-3}~M_\odot$. These values of mass and size are also allowed by the searches for lensing toward the LMC (Draine, 1998). Cloudlets can survive if the number of collisions is sufficiently small
and their stability is best explained if they are self-gravitating (Combes 2000). Concerning lensing, the average
surface density of a SGMC would be sub-critical, but the chaotic structure 
of the matter distribution may lead to a network of caustics with high local magnification.

In the present Letter I test the hypothesis that the magnitude anomaly in the quasar distribution can 
be explained by the lensing effect of an hypothetic SGMC in the Galactic halo.

\section{Micro lensing and cloud lens models}

\subsection{Introduction}

Lensing of distant galaxies and quasars shears and magnifies their images.
There are extensive reviews and books on lensing including Schneider, Ehlers \& Falco (1992), Mellier (1999), Schneider (2005, extensive references therein) and public algorithms.  
Fundamental theory, ray-tracing algorithms (Valle \& White 2003 and references therein), and observational techniques have been developed for two to three decades to 
infer cluster mass distributions (review by Gladders et al 2002), large scale structures and cosmological models (Soucail et al 2004; Wambsganss et al 1997, 2004; Sand et al 2005). 

Systematic studies
of micro-lensing were initiated in the early 90s, by Wambsganss (1990), who developed
ray-shooting simulations to show the magnification pattern of star fields. In the present paper we will mainly use the  public software LensTool (Jullo et al 2007) which includes strong and weak lens regimes.

The change in direction of a light ray can be written
$$ d {\bf a}  = -2~ {\rm grad}~ \phi ~d \chi$$
where $d \bf a$ is the bend angle, grad is the spatial gradient perpendicular
to the light path, $\phi$ is gravitational potential, and $\chi$ is the radial co-moving
coordinate. The distortion can be entirely described by the linearized lens mapping
through the Jacobi matrix $ \bf A$ of transform. This matrix is normally expressed with three
terms depending on derivatives of the the lensing potential: the convergence $\kappa$, the amplitude of 
the shear $\gamma$ and the rotation $\omega$. The convergence changes
the radius of a circular source, the shear changes its ellipticity, the rotation angle
is the phase of the shear. Rotation occurs in the case of multiple lens planes.

The convergence $\kappa = {1 \over 2} (\partial^2_1 + \partial^2_2)~ \phi$ at location $\theta$ is related to 
the projected surface density $\Sigma(\theta)$ at that location by
 
$$ \kappa (\theta) = {\Sigma(\theta) \over \Sigma_{\rm crit}} $$

in which the critical surface density  $\Sigma_{\rm crit} $ is given by 
$$ \Sigma_{\rm crit} = {c^2 \over {4 \pi G}} {D_s \over {D_l D_{ls}}} $$ 
where $D_s$, $D_l$ and $D_{ls}$ are respectively the distances observer to source, observer to lens, and 
lens to source. 

The components of the shear are given by 
$\gamma_1 = {1 \over 2} ( \partial^2_1 - \partial^2_2)~ \phi$ and $\gamma_2 = \partial_1 \partial_2~ \phi$. The shear amplitude is the module of its two components. 

The present Letter focusses on micro-lensing by numerous clumps; the convergence and the shear of any underlying smooth component will be ignored. 

\subsection{Preliminary test lens models}

The critical surface density of a smooth lens located in the Galactic halo on distant objects such that $D_s  \approx D_{ls} $ and assuming $D_l = \rm 15~ kpc$ is $\rm 10^5~ M_\odot pc^{-2}$, which leads to a mass of $\rm \sim 10^{14}~ M_\odot$ for a kpc scale SGMC, so a 3D filling factor of the order of $10^{-5}$ is required to fit in the halo. A series of cloud grid models of kpc size in depth, with individual clumps of mass $\rm  10^{-3}~ M_\odot$, 10 AU in diameter, are built with various filling factors. 

Cloud models are constructed as follows: an initial fits frame of $4000 \times 4000$ pixels with zero intensity including grids of the order of 10-100 pixels with common value 1 is created, so that typical separations of the order of 
$\rm \triangle \sim  10000~ AU$ are reached by assuming a physical pixel size of 
10 AU.  A series of layers is obtained by shifting randomly the coordinates of the initial frame which we co-add. We repeat this shift and 
add procedure on the first co-added layer to get a second generation, and repeat again the process up to several generations. For example a model with surface density $0.8 ~\Sigma_{\rm crit}$ and 1 kpc in depth, is 
built with $\rm 8^5 + 8^4~ layers$, from an initial frame with 55 non-zero pixels. This lens has $2 \times 10^6$ micro-lenses, some being combined on the same pixel. 

Lens mass models are obtained  by multiplying the final frames by the mass of an individual clump, and transformed into ascii files
assuming that luminous pixels are  \sl point masses. \rm Point masses are used for simplicity. A detailed lens analysis, in which self-gravitating gas clumps are described by polytropic models, can be found in Draine (1998). Lens simulations with LensTool are done on small sections of  the final frames including a sufficiently large number of point masses each. 

Three cases are considered: 

- $ \kappa = 0.2$ in order to compare our simulations with those in the literature, 

- $ \kappa = 0.09$ because the theoretical mean amplification calculated from
$$ < \mu > = {1 \over  {|(1 - \kappa )^2 - \gamma^2|}} $$
and assuming $\gamma = 0$ is that of Longo's effect $ < \mu > = 1.2$,  

- $ \kappa = 0.8$ and $ 0.9$ to explore cases with high magnification and dense caustics. Parameters of test lens models are given in Table~\ref{grid}. 

In the case $ \kappa = 0.2$, individual caustics can be identified, there are perturbations of single point caustics by other point masses giving complicated caustic structures comparable with those shown in Wambganss (1990). The average amplification value obtained by simulations is the same as the analytic value.

For $ \kappa = 0.09$, the magnification pattern differs from
the case $ \kappa = 0.2$: individual caustics can be identified but there is almost no 
perturbation of single clump caustics by other clumps presumably because the density is low. So there 
is no complicated caustic structure comparable to the previous case. The average amplification value 
is the same as the analytic value with a dispersion due to the small number of clumps per frame
section.  

For $ \kappa = 0.8$, the magnification pattern shows a high clustering of caustics. While the average magnification in simulations, $ < \mu > = 25$, is the same as the analytic value, large variations are detected with some locations showing huge amplification. We recover here the non-linear spatial distribution of amplifications mentioned by Wambganss (1990). 

All these $\rm kpc^3$ test models are by far too massive to be used  as elementary modules of a SuperGiant Cloud of the order of $\rm 400~ kpc^3$. Low filling factors are necessary.

A model with local surface density  $ \kappa = 0.8$ and a 2D filling factor of $f = 1.2/25 = 0.048$, namely a series of lenses with high amplification 25 diluted in a surface $\sim 20$ larger, was considered (test model E). This patchy model gives  the expected average magnification $ < \mu > = 1.2$ with a mass 2.3 times smaller than the case $ \kappa = 0.09$ with $f = 1$. The case $ \kappa = 0.9$ and a 2D filling factor of  $f = 1.2/100 = 0.012$ provides very high local magnification up to 100, an average  $ < \mu > = 1.2$ for a total mass reduced by a factor 9 (test model F). 

\begin{table*}
\renewcommand{\arraystretch}{0.9}
\centering 
\caption{\rm Main parameters for lens mass models. The columns indicate respectively: model identification, mean separation $\rm \triangle$ between clumps, 2 D filling factor $f$, average surface density in filled area $\rm \Sigma$ for a 1 kpc depth, average amplification $ < \mu> $, mass density $\rho$ per 
$\rm pc^3$,  mass $\rm M_1$ of a SGMC of 1 $\rm kpc^3$ . In models E \& F the density is given for the dense regions, and the 
magnification is given both for the dense regions and on average.}
\vspace*{0.05in}
\begin{tabular}{ c c c c c c c   } 
Model & $\rm \triangle ~ (AU)$  &   $f$  & $\rm \Sigma~ (M_\odot~pc^{-2})$ &  $\rm <\mu >$ & $\rm \rho ~(M_\odot pc^{-3})$ & $\rm M_1 ~(M_\odot)$ \\ \hline
    &            &            &                              &            &       & \\
A & $ 10000 $ & 1         &   $8 \times 10^3$ &  1.18   &  8    & $8 \times 10^{9}$ \\
B & $ 9856 $   & 1         & $9 \times 10^3$   &  1.21   &  9   & $9 \times 10^{9}$ \\
C & $  7590 $  & 1         & $2 \times 10^4$   &  1.53   & 20  & $2 \times 10^{10}$ \\
D & $ 4760$   &  1        &  $8 \times 10^4$  &  25      &  80    & $8 \times 10^{10}$ \\
E & $ 4760 $   &  0.048 & $8 \times 10^4$   &  25, 1.2     & 80    & $3.8\times 10^{9}$ \\
F & $ 4575 $   &  0.012 & $9 \times 10^4$   &  100, 1.2     & 90   & $10^{9}$ \\
\hline\hline
\label{grid}
\end{tabular}
\end{table*}

\subsection{Constrained lens models}

Our lens mass models must have densities lower than the limit of molecular cloud cores: if we assume a maximum mass $\rm \sim 10~ M_\odot$ within a scale length of $ \rm  L \sim 0.1~ pc$ above which molecular clouds would collapse (Kayama et al 1996), and extrapolate this density through Larson's (1981) scaling law $ \rm \rho \propto L^{-1.1}$, one gets an upper limit density $\rm \sim 350~M_\odot~pc^{-3}$ for a scale length of $\rm \sim 100~ pc$. Whereas our densities satisfy this criterium, a molecular cloud in equilibrium with scale $\rm 30 ~ pc$, close to our building block, would have a typical mass $\rm \sim 5 \times 10^{5}~M_\odot$ (Solomon et al 1987), and a density $\rm 18.5~M_\odot~pc^{-3}$ which is significantly below that of 
models D, E, \& F. 

Reducing the size of the grid cells quickly boosts the amplification. Therefore a solution may come from a fractal structure of Molecular Clouds with rather small cells satisfying the equilibrium condition, separated by less dense regions, all assembled in Giant Molecular Clouds 
up to a SGMC, as it was proposed by Pfenniger \& Combes (1994; hereafter PC) for cold dark $\rm H_2$ clouds. A cloud cell of  $\rm 5 \times 10^{5}~M_\odot$ for a
scale $\rm 30 ~ pc$ will be used as building block for fractal SGMC models below.

If masses obey a fractal scaling relation with dimension $D$ from mass $M_{\rm min}$ to $M_{\rm max}$ and scale $L_{\rm min}$ to $L_{\rm max}$ one
may write:
$$ {M \over M_{\rm min}} = \left( {L \over L_{\rm min}} \right)^D$$
up to $L_{\rm max}$. The mass at level $l$ consists in a number of clumps $N$ at level $l - 1$ related
to the scale by the relation:
$$ {L_{l-1} \over L_{l}} = N^{-1/D}$$
Clouds can survive if the number of collisions is sufficiently small, which according to
PC and Irwin, Widraw, \& English (2000) requires $1 \leq D \leq 2$. If $D \geq 2.33$, internal and external collisions are too numerous for maintaining a fragmented cloud. Two cloud models with two fractal dimensions are considered first: the case $ D = 1.64$ which
was suggested in the original PC paper, and the upper value $D = 2$. 

Extrapolating from $L_{\rm min} = 30~\rm pc$, $M_{min} = 5 \times 10^{5}~M_\odot$ to $L_{\rm max} = 5~\rm kpc$, and assuming a fractal dimension $ D = 1.64$, one gets a total mass $M_{\rm max} = 2.2 \times 10^{9}~M_\odot$ within  $\rm 125~kpc^3$, and $N \sim 4500$ cloud cells, i.e. $\sim 16$ building blocks on each axis. The surface density integrated over a column $L_{\rm max} = 5~\rm kpc$ of 16 blocks is $\rm \Sigma = 8.9 \times 10^3~ M_\odot~pc^{-2}$, which is the maximum. Parameters of this model are given in 
Table~\ref{gridb} (model G).
The median magnification derived from the probability $p(n)$ of a light ray to cross 
$ 0 \leq n  \leq 16$ cloud blocks, is found to be $\sum \mu (n) p(n)  = 1.03$, so half the surface
provides a magnification between 1.20 and 1.03 which is not sufficient to explain
Longo's effect. 

The case $D = 2$ has a different behavior (model H). Being more populated than for $D = 1.67$,
the densities are higher,  the median amplification is $\mu = 1.12$, and 25 \% of the surface has $1.18 \leq \mu \leq 1.48$. Regions with $\mu \geq 1.18$ do show networks of caustics and non gaussian amplification distributions. Model H appears at the limit of explaining Longo's effect with a 
mass of $4.1  \times 10^{10}~M_\odot$ within $ 28^\circ$. \rm 

\begin{table*}
\renewcommand{\arraystretch}{0.9}
\centering 
\caption{\rm Main parameters for fractal lens mass models. The columns indicate respectively: 
model identification,
lens plane distance $D_l$,
fractal dimension $D$,
mass of building block $M_{\rm min}$, 
scale of building block $L_{\rm min}$ in pc,
scale of SGMC $L$ in kpc, 
number $N$ of building blocks within $\rm 125~kpc^3$,
maximum integrated surface density $\rm \Sigma$ , 
median magnification $ \mu $, 
mass $\rm M_5$ of a SGMC within $\rm 125~kpc^3$.}
\vspace*{0.05in}
\begin{tabular}{  c c c c c c c c c c   }
 Model & $D_l$ &$D$ &  $M_{\rm min} \rm (M_\odot)$ & $L_{\rm min} \rm (pc)$ & $L \rm (kpc)$ & $N$ & $\rm \Sigma~ (M_\odot~pc^{-2})$ 
& $\rm \mu $ & $\rm M_5~(M_\odot)$ \\ \hline
 & & &      &                  &            &                              &                    &        &  \\
G & 16 & 1.64 &   $5 \times 10^5$   &  30  &  5 & 4 500    &  $8.9 \times 10^3$   & 1.03   & $2.2 \times 10^{9}$ \\
H & 16 & 2      & $5 \times 10^5$     & 30   & 5  & 27 750  &  $1.7 \times 10^4$ &     
1.12  & $1.4  \times 10^{10}$ \\
I & 20 & 1.8  & $5 \times 10^5$     & 30   & 5  & 10 106  & $1.2 \times 10^4$ & 1.12 &
$5  \times 10^{9}$ \\
\hline\hline
\label{gridb}
\end{tabular}
\end{table*}

\subsection{Shear}
The shear was not considered in our previous models. This is due to the fact that the shear is a term resulting from the mass outside the light beam, like a smooth galactic component to a star field. In our case however, cloud blocks are mass concentrations which may induce a shear component. A series of tests were done on sections of our grids using LensTool and were repeated with the MRLENS package (Starck et al 2006) which is dedicated to weak lensing.  There is considerable dispersion between grid sections with the same
average surface density. Shear is not detected on grid sections with a single block ($\rm 5 \times 10^5~M_\odot$), that is for grids with measured amplification $\mu =1.01$. Shear is significant in grid sections of average amplification $\mu = 1.1$ and above, but there is an important
spread $0.1 ~ \kappa \leq \gamma \leq \kappa$.  Assuming $\gamma = \kappa$, the
amplification would slightly change, so the same amplifications as for model with $D = 2$ would
be obtained with $D = 1.93$. In that case the mass within $\rm 125~kpc^3$ is $9.8 \times 10^{9}~M_\odot$, and that of a cloud  of 5 kpc in depth extending over $28^\circ$ is $2.9 \times 10^{10}~M_\odot$.
 
\subsection{Distance to the lens plane}
Locating the lens plane at an observer distance $D_l = \rm 20~ kpc$ rather than 16 kpc, reduces the critical surface density by $25 \%$ to $\Sigma_{\rm crit} = 7.5 \times 10^4~  M_\odot~pc^{-2}$ for infinitely distant objects.
For this critical surface density the same amplifications as in model H, are obtained for a
cloud with $D = 1.80$, a mass $5 \times 10^{9}~M_\odot$ within $\rm 125~kpc^3$, and
a total mass $1.5 \times 10^{10}~M_\odot$ for a SGMC of 5 kpc in depth and extending over $\rm 8.7 \times 8.7~ kpc^2$ (model I). 

\sl Increasing the depth to 8.6 kpc and using a dimension $D = 2$,  we obtain model J which provides a median amplification of  $\mu = 1.15$, with 14 \% of the surface amplified by more than 1.5. The total mass total within  $\rm 8.6 \times 8.6 \times 8.6~ kpc^3$ is $4.1 \times 10^{10}~M_\odot$. \rm

A single plane lens model has been assumed. If the cloud were spread over a large distance in depth, extending for example from 
$D_l = \rm 10~ kpc$ to 30 kpc, the critical surface density would vary by a factor 3 with
$D_l$, the distant part of the lens being more efficient than the nearby one. Over that distance, the cosmic-ray flux coming from the central part of the Galaxy would decrease by a factor 9, so if the cloud had a large depth, there would be a distance shift between the illuminated area and the average lens plane. The mass budget may be further optimized  by changing the depth of the SGMC and its fractal dimension, and consequently the probability of a light-ray to cross a sufficiently large number of cloud blocks.

\subsection{ H2 clumps and WIMPS}
$\rm H_2$ clumps are extremely dense compared with WIMPs clumps of the same mass, as predicted by N-body simulations (Diemand et al 
2005, Springel et al 2008). Let consider for example a WIMP clump seeded by an $\rm H_2$ clump of the same mass 
$\rm M = 10^{-3}~M_\odot$. The WIMP clump alone would have a radius of 1 000 AU 
in the absence of $\rm H_2$. Under the attraction of the central $\rm H_2$ clump, it will experience a first 
infall during which its density will increase by $10^6$ in $\rm \sim 1~ Myr$. If WIMPS are low mass neutralinos of $\rm m_\chi \sim 10~ GeV$, number density  
$\rm n_\chi$, and cross-section $\rm \sigma_{V=0} = 10^{-26}~ cm^3~ s^{-1}$ for annihilation (Cerdeno et al 2012), those concentrated in the clump will have significant annihilation, leading to a mass transform $\rm 1.4~10^{-7}~M_\odot~Myr^{-1}$ mainly into 
$p,~ p-$ and $\rm \gamma$. The cumulated mass loss of those neutralinos is $50 \%$ in $\rm \sim 6 ~Gyr$, thus an upper limit of $\sim 50~\%$ the clump mass may remain in WIMPs within 10 AU over cosmological ages. 

\section{Conclusion}
A luminosity excess of 0.2 mag was recently discovered in the SSDS survey of quasars. This effect which has a huge spatial extent of the order of $\rm 30^\circ$, is extremely well determined statistically, and coincides with a large and diffuse luminosity in $\gamma$ detected by the \sl Fermi Gamma-ray Space Telescope. \rm 

We make the working hypothesis that the detected $\gamma$ luminosity traces a SuperGiant Molecular Cloud illuminated by cosmic rays, at least partly, and test the hypothesis on whether such an cloud would be capable of explaining the quasar luminosity excess by micro-lensing.

Cloud models are built by using compact clumps of mass $10^{-3}~M_\odot$, 10 AU in size, as suggested by Extreme Scattering Events on quasars, typical molecular cloud building blocks of mass $5 \times 10^{5}~M_\odot$ for a scale of 30 pc, and a fractal structure of the SuperGiant Cloud with dimension $1.64 \leq D \leq 2$ up to a scale of 5-8.6 kpc,
and located at an observer-lens distance $D_l = \rm 16-20~kpc$.

A SGMC located at a distance $D_l = \rm 20~kpc$ from the observer with fractal dimension $D = 1.8$, total mass $1.5 \times 10^{10}~M_\odot$, 5 kpc in depth
and extending over 8.7 kpc \sl  do show amplifications of the order of Longo's effect in sparse regions. A SGMC with depth increased to 8.6 kpc and dimension $D = 2$ appears capable of explaining Longo's effect with a huge mass of $4.1 \times 10^{10}~M_\odot$. \rm

\vskip 0.5in
\vbox{
\noindent{ {\bf Acknowledgments} } \\
\noindent  

This paper is dedicated to the memory of Jean-Pierre Vigier.}

\end{document}